\documentclass[12pt,a4]{article}
\usepackage{amsmath}
\usepackage{bm}
\usepackage{amsfonts}
\usepackage{amssymb}
\usepackage{graphics}
\usepackage[normal]{caption2}
\usepackage{subfigure}
\usepackage{rotating}
\usepackage{citesort}
\def\be{\begin{equation}}
\def\ee{\end{equation}}
\def\ba{\begin{array}}
\def\ea{\end{array}}
\def\bea{\begin{eqnarray}}
\def\eea{\end{eqnarray}}

\begin{document}
\baselineskip 20pt \setlength\tabcolsep{2.5mm}
\renewcommand\arraystretch{1.5}
\setlength{\abovecaptionskip}{0.1cm}
\setlength{\belowcaptionskip}{0.5cm}
\begin{center} {\large\bf On the density and temperature of neutron-rich systems at the energy of vanishing flow in heavy-ion collisions}\\
\vspace*{0.4cm}
{\bf Sakshi Gautam} \footnote{Email:~sakshigautm@gmail.com}\\
{\it  Department of Physics, Panjab University, Chandigarh -160
014, India.\\}
\end{center}
We study nuclear dynamics at the the energy of vanishing flow for
neutron-rich systems. In particular, we shall study the collision
rate, density and temperature reached in a heavy-ion reaction with
neutron-rich systems. We shall also study the mass dependence of
these quantities. Our results indicate nearly mass independent
nature for  the density reached whereas a significant mass
dependence exists for the temperature of neutron-rich systems.


\newpage
\baselineskip 20pt
\section{Introduction}
The collective transverse in-plane flow has been used extensively
over the past three decades to study the properties of hot and
dense nuclear matter., i.e., the nuclear matter equation of state
(EOS) and in-medium nucleon-nucleon cross section \cite{sch}. It
has been reported to be highly sensitive to the above mentioned
properties and also to the entrance channel parameters like
incident energy, colliding geometry and system size
\cite{ogli89,andro03,luka05,zhang06,luka08}. The energy dependence
of flow led to its disappearance at a particular incident energy
called energy of vanishing flow (EVF) or balance energy
(E$_{bal}$) \cite{krof89}. A large number of theoretical studies
have been carried out in the past studying the sensitivity of
E$_{bal}$ to the system size and colliding geometry
\cite{mag00,sood1,mag100}.
\par
Role of isospin degree of freedom in collective transverse
in-plane flow and its disappearance has also been a matter of
great interest for the past decade \cite{pak97,li}. Isospin degree
of freedom plays its role in determining the nuclear equation of
state of asymmetric nuclear matter. The availability of
radioactive ion beams (RIBs) \cite{rib1,rib2} around the world
helps in carrying out the studies on the matter lying far away
from the stability line. A number of studies have been carried out
in the recent past to see the role of isospin degree of freedom in
collective flow and its disappearance \cite{pak97,li,gaum1}. In
Ref. \cite{gaum2} author and others studied the isospin effects in
E$_{bal}$ at all the colliding geometries. A very few studies have
been carried out to study other related phenomena at E$_{bal}$ of
the neutron-rich systems. In Ref. \cite{sood2} Sood and Puri have
presented a complete study of the nuclear dynamics at E$_{bal}$
for stable systems. The study includes participant-spectator
matter, density and temperature reached in a heavy-ion reaction at
E$_{bal}$. Motivated by this, author and others presented a study
of participant-spectator matter of neutron-rich systems at
E$_{bal}$ in Ref. \cite{gaum3}. The study revealed a similar
behaviour of participant-spectator matter for neutron-rich systems
as for the stable systems. Another important quantity which
reflects the dynamics in a heavy-ion collision is the density and
temperature reached in a reaction. In the present paper, we study
the density and temperature reached in heavy-ion reactions of
neutron-rich matter at E$_{bal}$. We also aim to see the role of
isospin degree of freedom in the density and temperature reached
in the reactions of neutron-rich systems and to see if the
behaviour of these properties at balance energy differs from that
for systems lie close to the stability line.
 \par
The present study is carried out within the framework of
isospin-dependent quantum molecular dynamics (IQMD) model
\cite{hart98}. Section 2 describes the model in brief. Section 3
explains the results and gives our discussion, and Sec. 4
summarizes the results.
 \par
 \section{The model}
 The IQMD model \cite{hart98} which is the extension of quantum molecular dynamics (QMD) \cite{aichqmd}
 model treats different charge states of
nucleons, deltas, and pions explicitly, as inherited from the
Vlasov-Uehling-Uhlenbeck (VUU) model. The IQMD model has been used
successfully for the analysis of a large number of observables
from low to relativistic energies. Puri and coworkers have
demonstrated that QMD, IQMD carries essential physics needed to
demonstrate the various phenomena such as collective flow,
multifragmentation and particle production \cite{dhawan,batko}.
The isospin degree of freedom enters into the calculations via
symmetry potential, cross sections, and Coulomb interaction.
 \par
 In this model, baryons are represented by Gaussian-shaped density distributions

\begin{equation}
f_{i}(\vec{r},\vec{p},t) =
\frac{1}{\pi^{2}\hbar^{2}}\exp(-[\vec{r}-\vec{r_{i}}(t)]^{2}\frac{1}{2L})
\times \exp(-[\vec{p}- \vec{p_{i}}(t)]^{2}\frac{2L}{\hbar^{2}})
 \end{equation}

 Nucleons are initialized in a sphere with radius R = 1.12 A$^{1/3}$ fm, in accordance with liquid-drop model.
 Each nucleon occupies a volume of \emph{h$^{3}$}, so that phase space is uniformly filled.
 The initial momenta are randomly chosen between 0 and Fermi momentum ($\vec{p}$$_{F}$).
 The nucleons of the target and projectile interact by two- and three-body\textrm{ Skyrme} forces, \textrm{Yukawa} potential, and \textrm{Coulomb} interactions. In addition to the use of explicit charge states of all baryons
and mesons, a symmetry potential between protons and neutrons
 corresponding to the Bethe-Weizs\"acker mass formula has been included. The hadrons propagate using Hamilton equations of motion:

\begin {eqnarray}
\frac{d\vec{{r_{i}}}}{dt} = \frac{d\langle H
\rangle}{d\vec{p_{i}}};& & \frac{d\vec{p_{i}}}{dt} = -
\frac{d\langle H \rangle}{d\vec{r_{i}}}
\end {eqnarray}

 with

\begin {eqnarray}
\langle H\rangle& =&\langle T\rangle+\langle V \rangle
\nonumber\\
& =& \sum_{i}\frac{p^{2}_{i}}{2m_{i}} + \sum_{i}\sum_{j>i}\int
f_{i}(\vec{r},\vec{p},t)V^{\textrm{ij}}(\vec{r}~',\vec{r})
 \nonumber\\
& & \times f_{j}(\vec{r}~',\vec{p}~',t) d\vec{r}~ d\vec{r}~'~
d\vec{p}~ d\vec{p}~'.
\end {eqnarray}

 The baryon potential V$^{\textrm{ij}}$ in the above relation, reads as

 \begin {eqnarray}
  \nonumber V^{\textrm{ij}}(\vec{r}~'-\vec{r})& =&V^{\textrm{ij}}_{\textrm{Skyrme}} + V^{\textrm{ij}}_{\textrm{Yukawa}} +
  V^{\textrm{ij}}_{\textrm{Coul}} + V^{\textrm{ij}}_{\textrm{sym}}
    \nonumber\\
   & =& [t_{1}\delta(\vec{r}~'-\vec{r})+t_{2}\delta(\vec{r}~'-\vec{r})\rho^{\gamma-1}(\frac{\vec{r}~'+\vec{r}}{2})]
   \nonumber\\
   &  & +t_{3}\frac{\exp(|(\vec{r}~'-\vec{r})|/\mu)}{(|(\vec{r}~'-\vec{r})|/\mu)}+
    \frac{Z_{i}Z_{j}e^{2}}{|(\vec{r}~'-\vec{r})|}
   \nonumber \\
      &  & +t_{4}\frac{1}{\varrho_{0}}T_{\textrm{3i}}T_{\textrm{3j}}\delta(\vec{r_{i}}~'-\vec{r_{j}}).
 \end {eqnarray}

Here \emph{Z$_{i}$} and \emph{Z$_{j}$} denote the charges of
\emph{ith} and \emph{jth} baryon, and \emph{T$_{3i}$} and
\emph{T$_{3j}$} are their respective \emph{T$_{3}$} components
(i.e., $1/2$ for protons and $-1/2$ for neutrons). The
parameters\emph{ $\mu$} and \emph{t$_{1}$,...,t$_{4}$} are
adjusted to the real part of the nucleonic optical potential.
 For the density dependence of  the nucleon optical potential, standard \textrm{Skyrme} type parametrization is employed.
We use a soft equation of state along with the standard isospin-
and energy-dependent cross section reduced by
  20$\%$, i.e. $\sigma$ = 0.8 $\sigma_{nn}^{free}$. In a recent study, Gautam \emph{et al}. \cite{gaum1}
  has confronted the theoretical calculations of IQMD with the data of $^{58}Ni+^{58}Ni$ and $^{58}Fe+^{58}Fe$ \cite{pak97}.
  The results with the soft EOS (along with the
momentum-dependent interactions) and above choice of cross section
are in good agreement with the data at all colliding geometries.
The details about the elastic and inelastic cross sections for
proton-proton and proton-neutron collisions can be found in
\cite{hart98,cug}. The cross sections for neutron-neutron
collisions are assumed to be equal to the proton-proton cross
sections. Two particles collide if their minimum distance\emph{ d}
fulfills
\begin {equation}
 d \leq d_{0} = \sqrt{\frac{\sigma_{tot}}{\pi}},   \sigma_{tot} =
 \sigma(\sqrt{s}, type),
\end {equation}
where 'type' denotes the ingoing collision partners (N-N....).
Explicit Pauli blocking is also included; i.e. Pauli blocking of
the neutrons and protons is treated separately. We assume that
each nucleon occupies a sphere in coordinate and momentum space.
This trick yields the same Pauli blocking ratio as an exact
calculation of the overlap of the Gaussians will yield. We
calculate the fractions P$_{1}$ and P$_{2}$ of final phase space
for each of the two scattering partners that are already occupied
by other nucleons with the same isospin as that of scattered ones.
The collision is blocked with the probability
\begin {equation}
 P_{block} = 1-[1 - min(P_{1},1)][1 - min(P_{2},1)],
\end {equation}
and, correspondingly is allowed with the probability 1 -
P$_{block}$. For a nucleus in its ground state, we obtain an
averaged blocking probability $\langle P_{block}\rangle$ = 0.96.
Whenever an attempted collision is blocked, the scattering
partners maintain the original momenta prior to scattering.
\par
 \section{Results and discussion}
We simulate the reactions of Ca+Ca, Ni+Ni, Zr+Zr, Sn+Sn, and Xe+Xe
series having N/Z = 1.0, 1.6 and 2.0. In particular, we simulate
the reactions of $^{40}$Ca+$^{40}$Ca (105), $^{52}$Ca+$^{52}$Ca
(85), $^{60}$Ca+$^{60}$Ca (73); $^{58}$Ni+$^{58}$Ni (98),
$^{72}$Ni+$^{72}$Ni (82), $^{84}$Ni+$^{84}$Ni (72);
$^{81}$Zr+$^{81}$Zr (86), $^{104}$Zr+$^{104}$Zr (74),
$^{120}$Zr+$^{120}$Zr (67); $^{100}$Sn+$^{100}$Sn (82),
$^{129}$Sn+$^{129}$Sn (72), $^{150}$Sn+$^{150}$Sn (64) and
$^{110}$Xe+$^{110}$Xe (76), $^{140}$Xe+$^{140}$Xe (68) and
$^{162}$Xe+$^{162}$Xe (61) at an impact parameter of
b/b$_{\textrm{max}}$ = 0.2-0.4  and at the incident energies equal
to
 balance energy. The values in the brackets represent
the balance energies for the systems. The reactions are followed
till the transverse in-plane flow saturates. It is worth
mentioning here that the saturation time varies with the mass of
the system. It has been shown in Ref. \cite{sood4} that the
transverse in-plane flow in lighter colliding nuclei saturates
earlier compared to heavy colliding nuclei. Saturation time is
about 100 (150 fm/c) in lighter (heavy) colliding nuclei in the
present energy domain. We use the quantity "\textit{directed
transverse momentum $\langle p_{x}^{dir}\rangle$}" to define the
nuclear transverse in-plane flow, which is defined as
\cite{hart98,aichqmd,leh}

\begin {equation}
\langle{p_{x}^{dir}}\rangle = \frac{1} {A}\sum_{i=1}^{A}{sign\{
{y(i)}\} p_{x}(i)},
\end {equation}
where $y(i)$ and $p_{x}$(i) are, respectively, the rapidity
(calculated in the center of mass system) and the momentum of the
$i^{th}$ particle. The rapidity is defined as

\begin {equation}
Y(i)= \frac{1}{2}\ln\frac{{\vec{E}}(i)+{\vec{p}}_{z}(i)}
{{\vec{E}}(i)-{\vec{p}}_{z}(i)},
\end {equation}

where $\vec{E}(i)$ and $\vec{p_{z}}(i)$ are, respectively, the
energy and longitudinal momentum of the $i^{th}$ particle. In this
definition, all the rapidity bins are taken into account.

\begin{figure}[!t] \centering
 \vskip -1cm
\includegraphics[width=10cm]{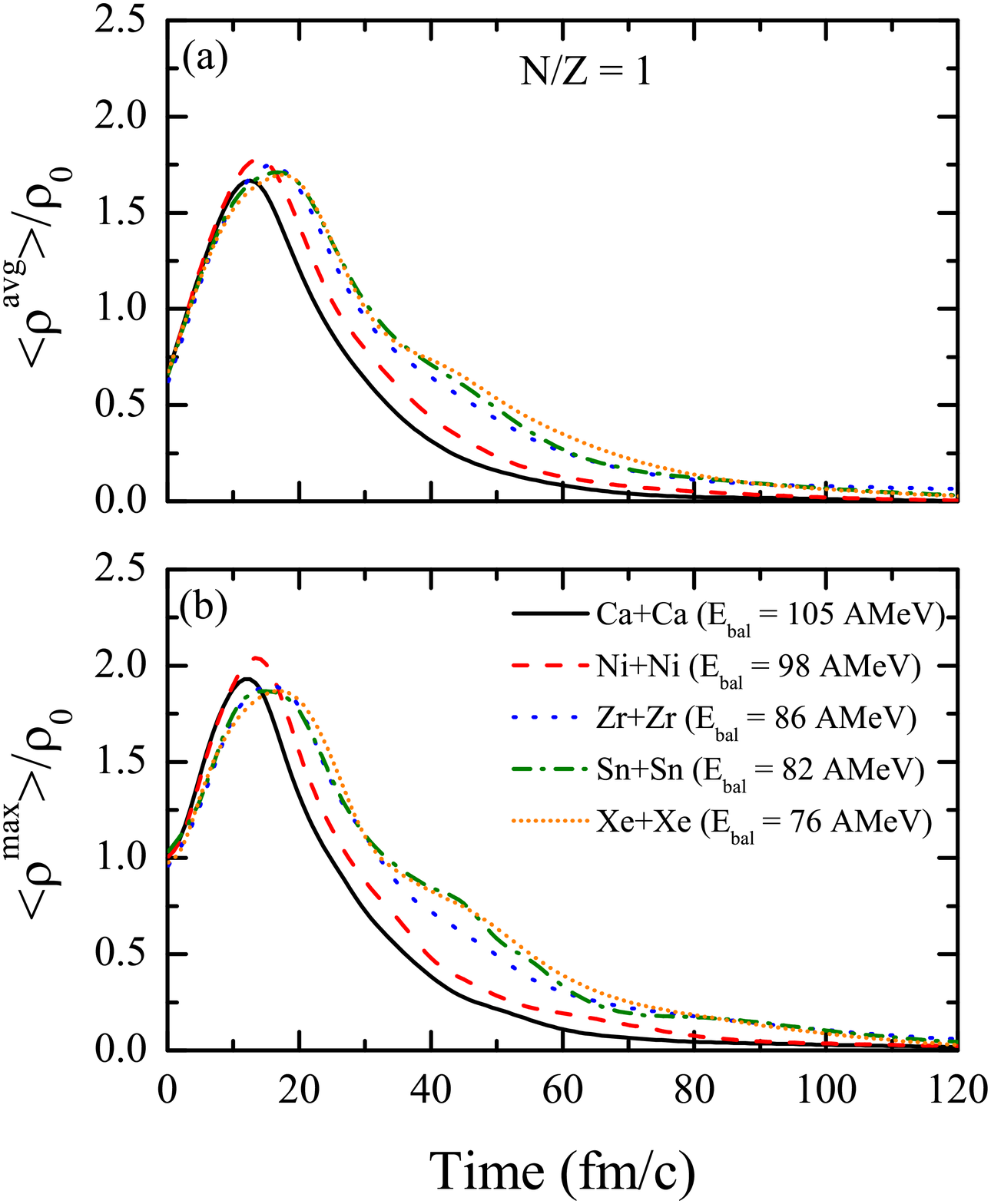}
\caption{(Color online) The time evolution of spectator matter
(left panels) and participant matter (right panels) for systems
having N/Z = 1.0, 1.6 and 2.0. Lines are explained in the
text.}\label{fig1}
\end{figure}

\begin{figure}[!t] \centering
 \vskip 1cm
\includegraphics[angle=0,width=10cm]{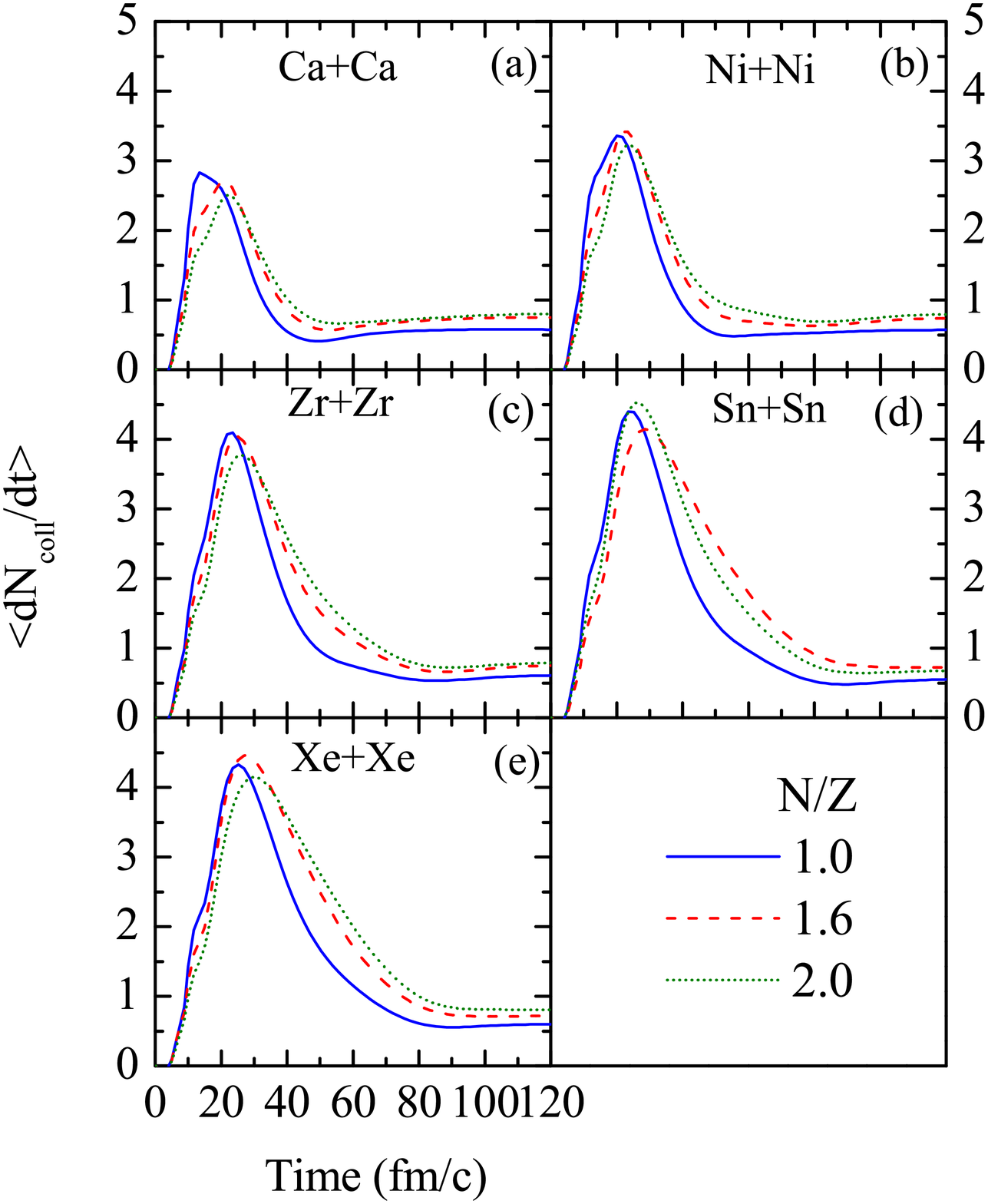}
 \vskip -0cm \caption{ (Color online) The N/Z dependence of participant and spectator matter. Symbols are explained in the text.} \label{fig2}
\end{figure}

\begin{figure}[!t] \centering
\vskip 0.5cm
\includegraphics[angle=0,width=8cm]{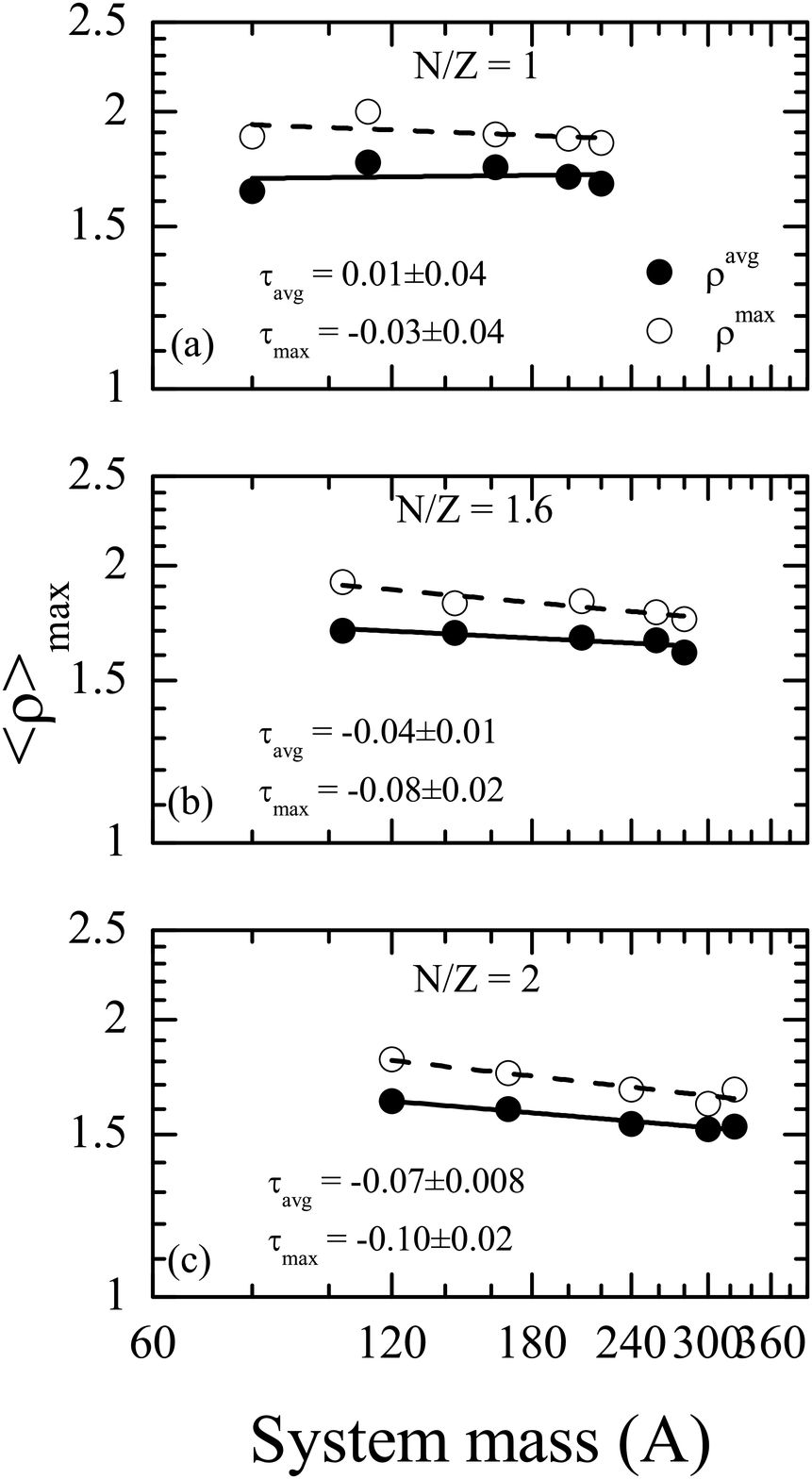}
\vskip 0.5cm \caption{ The system size dependence of participant
and spectator matter for different N/Z ratios. Various symbols are
explained in the text.}\label{fig3}
\end{figure}
\par
In fig. 1(a), we display the time evolution of average density
($\rho^{avg}/\rho_{0}$) whereas fig. 1(b) displays the time
evolution of maximum density ($\rho^{max}/\rho_{0}$) for the
systems having N/Z = 1.0, i.e, we display the reactions of
$^{40}$Ca+$^{40}$Ca, $^{58}$Ni+$^{58}$Ni, $^{81}$Zr+$^{81}$Zr,
$^{100}$Sn+$^{100}$Sn, and $^{110}$Xe+$^{110}$Xe at energy equal
to balance energy. Lines represent the different systems. Solid,
dashed, dotted, dash-dotted, and short dotted lines represent the
reactions of Ca+Ca, Ni+Ni, Zr+Zr, Sn+Sn, and Xe+Xe, respectively.
From figure, we find that maximal value of $\rho^{avg}/\rho_{0}$
is higher for lighter systems as compared to the heavier ones.
Moreover, the density profile is more extended in heavier systems
indicating that the reaction finishes later in heavier systems.
This is because of the fact that the heavier reaction occurs at
low incident energy. Also the $\rho^{avg}/\rho_{0}$ and
$\rho^{max}/\rho_{0}$ are nearly same for heavier systems but
differ for lighter systems as in Ref. Further, the maximum and
average densities are comparable for medium and heavy mass systems
indicating that the dense matter is formed widely and uniformly in
the central zone of the reaction. On the other hand, the
substantial difference in two densities for the lighter colliding
nuclei has been explained in Ref. \cite{sood2} and indicates the
non-homogeneous nature of dense matter. It is worth mentioning
that collective flow saturates at higher densities whereas
multifragmentation occurs at sub-density zone. Other phenomena
such as fusion-fission are very low density phenomena
\cite{puril}.

\begin{figure}[!t] \centering
 \vskip 1cm
\includegraphics[angle=0,width=12cm]{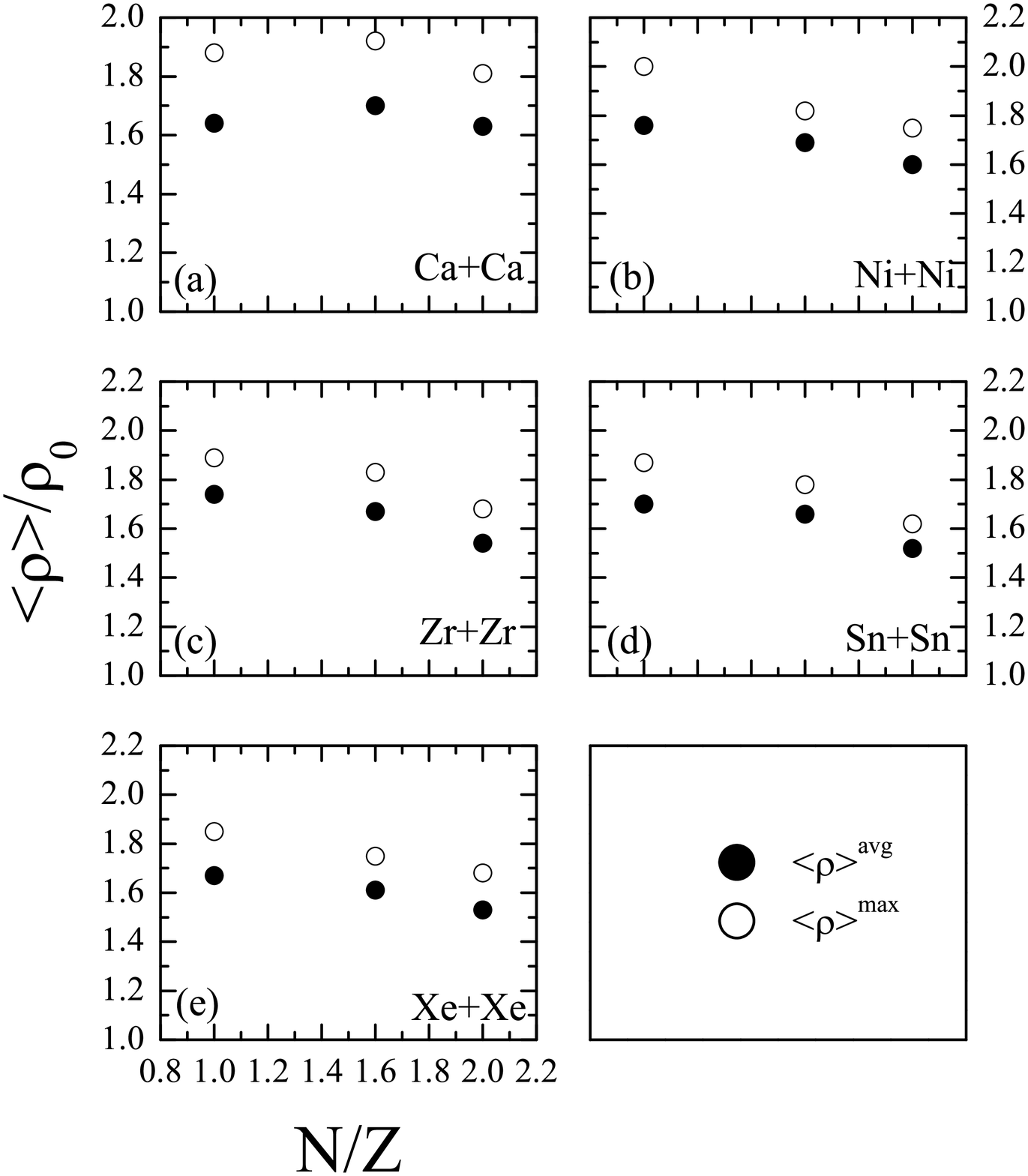}
 \vskip -0cm \caption{ The N/Z dependence of participant and spectator matter. Symbols are explained in the text.} \label{fig2}
\end{figure}

\par
The quantity which reflects the density achieved in a reaction is
the collision rate.  In fig. 2, we display the time evolution of
the collision rate for various systems having N/Z = 1.0, 1.6 and
2.0. Solid, dashed and dotted lines corresponds to systems having
N/Z = 1.0, 1.6 and 2.0, respectively. From figure, we see that
collision rate first increases with time, reaches maximum at
around 20-40 fm/c (which is the high dense phase of the reaction)
and then finally decreases and becomes constant at around 80 fm/c.
We also find that the maximum value of the collision rate also
increases with the system mass. Moreover, the effect of N/Z ratio
on the collision rate is very less.
\par
In fig. 3 we display the system size dependence of maximal value
of the maximum ($\rho^{max}$) and average density ($\rho^{avg}$)
for the systems having N/Z = 1.0, 1.6 and 2.0. We see that the
maximal value of $\rho^{max}$  and $\rho^{avg}$ follows a power
law behaviour proportional to A$^{\tau}$. The power law factor is
0.01$\pm$ 0.04 (-0.03$\pm$ 0.04), -0.04$\pm$ 0.01 (-0.08$\pm$
0.02), and -0.07$\pm$ 0.008 (-0.10$\pm$ 0.02) for $\rho^{avg}$ (
$\rho^{max}$) for systems with N/Z = 1.0, 1.6 and 2.0,
respectively. It shows that the dependence of maximal value of
$\rho^{avg}$  and  $\rho^{max}$ is very weak on the system size
for all the N/Z ratios. This was also predicted in Ref.cite{sood2}
for stable systems.
 \begin{figure}[!t] \centering
 \vskip -1cm
\includegraphics[width=8cm]{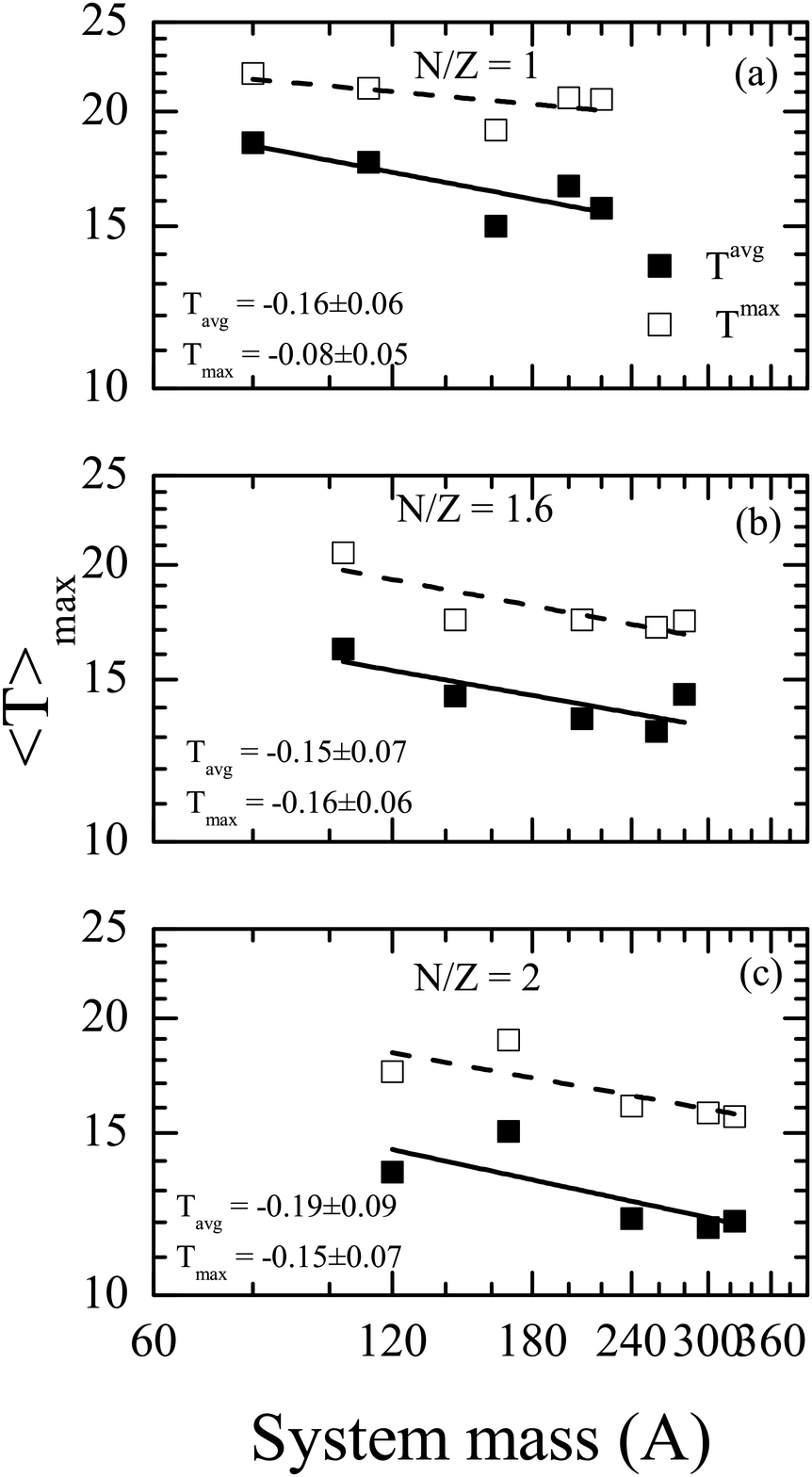}
\caption{ The system size dependence of anisotropy ratio for
various N/Z ratios.}\label{fig5}
\end{figure}


 \begin{figure}[!t] \centering
 \vskip -1cm
\includegraphics[width=12cm]{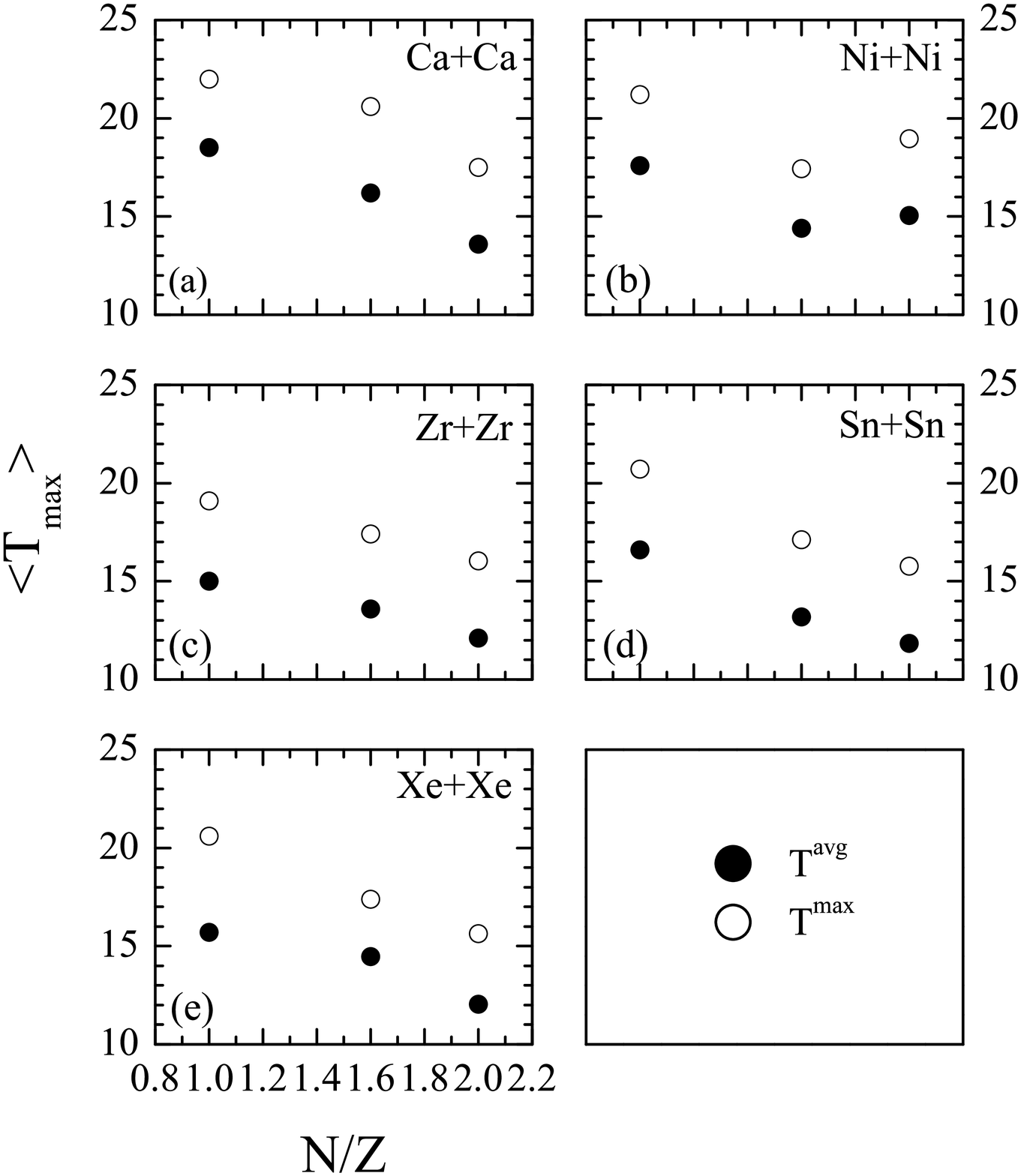}
\caption{The system size dependence of anisotropy ratio for
various N/Z ratios.}\label{fig5}
\end{figure}


\par
In fig. 4 we display the N/Z dependence of maximal value of
$\rho^{avg}$ and $\rho^{max}$. Solid (open) symbols display the
results for  $\rho^{avg}$ ( $\rho^{max}$). From figure we see that
both  $\rho^{avg}$  and  $\rho^{avg}$  decreases slightly with N/Z
of the system for all the system masses. A slight exception to
this is there for the lighter mass of Ca+Ca.
\par
The another associated quantity linked with the dense matter is
the temperature. In principle, a true temperature can be defined
only for a thermalized and equilibrated matter. Since in heavy-ion
collisions the matter is non-equilibrated, one can not talk of
``temperature''. One can, however, look in terms of the local
environment only. In our present case, we follow the description
of the temperature given in Refs. \cite{khoa921,khoa922}.
 In the present case, extraction of the temperature {\it T} is based on the local
density approximation, i.e., one deduces the temperature in a
volume element surrounding the position of each particle at a
given time step \cite{khoa921,khoa922}. Here, we postulate that
each local volume element of nuclear matter in coordinate space
and time has some ``temperature'' defined by the diffused edge of
the deformed Fermi distribution consisting of two colliding Fermi
spheres, which is typical for a nonequilibrium momentum
distribution in heavy-ion collisions.

In this formalism (dubbed the hot Thomas-Fermi approach
\cite{khoa921}), one determines extensive quantities like the
density and kinetic energy as well as entropy with the help of
momentum distributions at a given temperature. Using this
formalism, we also extracted the average and maximum temperature
within a central sphere of 2 fm
radius as described in the case of density.\\
In fig. 5 we display the maximal value of $<T^{avg}>$ and
$<T^{max}>$ as a function of the composite mass of the system.
From figure, we see that $<T^{avg}>$  and $<T^{max}>$ follows a
power law behaviour with system mass for all the N/Z ratios. The
power law factor is -0.16$\pm$ 0.06 (-0.08$\pm$ 0.05), -0.15$\pm$
0.07 (-0.16$\pm$ 0.06), and -0.19$\pm$ 0.09 (-0.15$\pm$ 0.07) for
for $<T^{avg}>$ ($<T^{max}>$) for systems having N/Z = 1.0, 1.6
and 2.0, respectively. Similar power law behaviour was also
predicted in Ref. \cite{sood2} for stable systems (N/Z $\simeq$
1). This system size dependence of temperature is sharp in
contrast with the density. This is because of the fact that the
temperature depends on the kinetic energy of the system.
 \par
In fig. 6 we display the N/Z dependence of maximal value of
$<T^{avg}>$ and $<T^{max}>$ for various system masses. Solid
(open) symbols represent the maximal value of $<T^{avg}>$
($<T^{max}>$). From figure, we see that for all the system masses,
$<T^{avg}>$ and $<T^{max}>$ decreases with N/Z of the system.

\par
\section{Summary}
 We studied the collision rate, density and temperature reached in
 reactions of neutron-rich systems at energy of vanishing flow.
 Our results pointed the similar behvaiour for neutron-rich
 systems as for the stable systems. We also investigated the mass
 dependence of these quantities. We found a very weak mass
 dependence of the density although the temperature follows a
 significant mass dependence.
 \par
This work has been supported by a grant from Centre of Scientific
and Industrial Research (CSIR), Govt. of India. Author is thankful
to Profs. J. Aichelin and R. K. Puri for enlightening discussions
on the present work.

\end{document}